\documentclass[aps,pre,twocolumn,superscriptaddress,showpacs]{revtex4-1}

\usepackage{amssymb}
\usepackage{color}
\usepackage{graphicx}
\usepackage{amsmath}
\usepackage{mathrsfs}
\usepackage{times}
\usepackage{subeqnarray}
\usepackage{cases}
\usepackage{bm}

\begin{document}

\title{Network synchronization with periodic coupling}

\author{Sansan Li}
\affiliation{School of Physics and Information Technology, Shaanxi Normal University, Xi'an 710062, China}

\author{Na Sun}
\affiliation{School of Physics and Information Technology, Shaanxi Normal University, Xi'an 710062, China}

\author{Li Chen}
\affiliation{School of Physics and Information Technology, Shaanxi Normal University, Xi'an 710062, China}

\author{Xingang Wang}
\email[Email address: ]{wangxg@snnu.edu.cn}
\affiliation{School of Physics and Information Technology, Shaanxi Normal University, Xi'an 710062, China}
%\affiliation{Department of Physics, Shaanxi Normal University, Xi'an 710062, China}

\begin{abstract}
The synchronization behavior of networked chaotic oscillators with periodic coupling is investigated. It is observed in simulations that the network synchronizability could be significantly influenced by tuning the coupling frequency, even making the network alternated between the synchronous and non-synchronous states. By the method of master stability function, we conduct a detailed analysis on the influence of coupling frequency on network synchronizability, and find that the network synchronizability is maximized at some characteristic frequencies comparable to the intrinsic frequency of the local dynamics. Moremore, it is found that as the amplitude of the coupling increases, the characteristic frequencies are gradually decreased. By the technique of finite-time Lyapunov exponent, we investigate further the mechanism for the maximized synchronizability, and find that at the characteristic frequencies the power spectrum of the finite-time Lyapunov exponent is abruptly changed from the localized to broad distributions. When this feature is absent or not prominent, the network synchronizability is less influenced by the periodic coupling. Our study shows the efficiency of finite-time Lyapunov exponent in exploring the synchronization behavior of temporally coupled oscillators, and sheds lights on the interplay between the system dynamics and structure in the general temporal networks.
\end{abstract}

%\date{\today }
%\pacs{05.45.Xt, 89.75.Hc}
\maketitle

\section{Introduction}

As a universal concept in nonlinear science, synchronization has been extensively studied over the past decades~\cite{SYNBOOK:Kuramoto,SYNBOOK:Pikovsky,SYNBOOK:Strogatz}. Recently, with the blooming of network science, the synchronization behavior of complex networks has attracted considerable attention~\cite{SYNREV:Boccaletti,SYNREV:Arenas}, in which one of the central tasks is to explore the interplay between the network dynamics and structure. For networks of linearly coupled identical oscillators, the conditions for synchronization can be well analyzed by the method of maser stability function (MSF)~\cite{MSF:Pecora,MSF:HG}, which suggests that for the given nodal dynamics and coupling function, the network synchronizability is determined by the eigenvalues calculated from the network coupling matrix~\cite{MSF:Barahona}. The MSF method provides a powerful tool for investigating network synchronization, based on which the influence of network structure on synchronizability has been well explored ~\cite{MSF:Barahona,NETSYN:CKHU,NETSYN:Nishikawa,NETSYN:Motter2005,NETSYN:WXG2007}. For instance, it has been shown that by introducing a few of random shortcuts onto a regular network, the network synchronizability can be significantly increased~\cite{MSF:Barahona,NETSYN:CKHU,NETSYN:Nishikawa}; and, by weighting the coupling weights according to the node degrees, the synchronizability of scale-free networks could be higher to that of random networks~\cite{NETSYN:Motter2005,NETSYN:WXG2007}. In these studies, the network structure is generally assumed as static, i.e., the network coupling matrix does not change with time~\cite{SYNREV:Boccaletti,SYNREV:Arenas}.

Realistic networks are typically non-static~\cite{TemNet:Holme,TemNet:LiAming}, e.g., the infrastructure networks are expanding in size, the connectivities of social networks are getting denser, the strengths of synapses in neuronal networks are modified according to external stimuli, the links in metabolic networks are activated only during specific tasks, to name just a few. The non-static feature of complex networks calls for the study of evolutionary networks, in which the network dynamics and structure (including network size, connectivity, link weights, etc) are mutually influenced and evolving with time together~\cite{SND:2002,SG:2007}. In exploring evolutionary networks, a key question is about the time scales of the following two dynamics~\cite{SND:2002}: one for the collective behavior of the networked nodes, $T_c$, and the other for the dynamical evolution of the network structure, $T_e$. When the time scales are separable~\cite{ZCS:2006,FW:2011,MWFDL:2011}, the network structure can be treated as either static ($T_c\ll T_e$) or as globally coupled by the time-average technique ($T_c\gg T_e$). In such cases, the network dynamics can still be analyzed by the conventional approaches. Challenges arise when the time scales are comparable ($T_c\sim T_e$)~\cite{SND:2002}. In this case, the network dynamics and structure evolution are strongly coupled, resulting in many intriguing phenomena~\cite{MOTTER:2005,Holme:2006,IBS:2010,WYF:2016}.

As an important approach to exploring the dynamics of evolutionary networks, synchronization in temporal networks has been studied in recent years~\cite{Belykh:2004,Boccaletti:2006,CM:2007,SO:2008,FM:2008PRL,CL:2009PRE,CL:2010EPJB,Porfiri:2012,Hasler:SIADS2,VK:2014,TIMME:PRL,ZhouJ:2016,FN:2016Chaos,AB:NODY,Golovneva:2017}. One important feature of temporal network is that the properties of the network links are timely varying~\cite{TemNet:Holme}, saying, for example, the blinking of the network structure~\cite{Belykh:2004}, the on-off switching of the links~\cite{CL:2009PRE}, the variation of the link weight and coupling function~\cite{ZCS:2006}. Due to the time-dependent links, some new synchronization phenomena have been observed in temporal networks and, to analyze these phenomena, a set of new methods and techniques have been developed. One typical model of temporal network is the blinking network~\cite{Belykh:2004,Porfiri:2012,Hasler:SIADS2,FM:2008PRL,VK:2014,FN:2016Chaos}, which is generally described as a fast, random switching of the shortcut links on a small-world network. Comparing to the static case, it is shown that the synchronizability of blinking network is significantly improved~\cite{Belykh:2004,Porfiri:2012,Hasler:SIADS2}. In blinking network, an important requirement is that the network structure should be switched at a very fast speed, i.e., $T_e\ll T_c$, with $T_e$ the time interval for the network to be staying on a specific structure and $T_c$ the characteristic time for network synchronization. The lower bound of the switching speed has been analyzed in Ref. \cite{Hasler:SIADS2}, which shows that beyond this bound the blinking network can be well represented by a time-averaged, static coupling matrix. Synchronization of networks with slowly switching structures are also tractable~\cite{SO:2008,ZhouJ:2016}, due to the separated time scales ($T_e\gg T_c$). In this case, the network dynamics can be treated as evolving on a sequence of static networks, and the network synchronizability can be analyzed by the MSF method or its generalized forms.

Interesting phenomena occur when the network structure is evolving at the moderate speeds, i.e., when $T_e\sim T_c$. By the strategy of on-off coupling, Chen \emph{et al.} studied the impact of switching frequency on network synchronizability~\cite{CL:2009PRE}, and found that when the time scale of the switching is comparable to that of the nodal dynamics, the stable synchronization region in the parameter space of the coupling strength could be significantly modified. In particular, at some specific frequencies of the switching, the upper bound of the stable region can be removed, making the network synchronizable over an infinite range in the parameter space. Similar phenomena have been also observed in transiently uncoupled chaotic oscillators~\cite{TIMME:PRL}. In this scheme, the coupling is activated only when the trajectory of the driven oscillator enters a ``clipping" region in the state space. It is shown that~\cite{TIMME:PRL}, by a suitable choice of the column of the clipping region, the upper bound of the stable synchronization region can also be removed. More recently, by the strategy of switching coupling, Buscarino \emph{et al.} studied the synchronization of two mutually coupled chaotic oscillators and found that, whereas the oscillators are not synchronizable by constant couplings, a switching between two such couplings at a moderate frequency could generate synchronization~\cite{AB:NODY}. Interestingly, it is also found that when the switching frequency is close to the intrinsic frequency of the oscillators, the oscillators are desynchronized. The desynchronization is attributed to the phase-locking between the periodic switching and the nodal dynamics, which results in the deformation of the oscillator attractors~\cite{AB:NODY}.

Despite the progresses mentioned above, synchronization of networked oscillators with time-dependent couplings remains an open question. In particular, the mechanism for the enhanced synchronization as induced by time-dependent coupling is still not very clear~\cite{CL:2009PRE,TIMME:PRL,AB:NODY}. In the present work, by the kernel of sinusoidal function, we revisit the synchronization of networked oscillators with time-dependent coupling, focusing on the temporal (local) instability of the synchronous manifold. In specific, by tuning the oscillating frequency of the couplings around the intrinsic frequency of the oscillators, we investigate the variation of the network synchronizability with respect to the coupling frequency, and, by the technique of finite-time Lyapunov exponent (FLE), analyze the temporal response of FLE to the periodic coupling. Our main finding is that network synchronization is optimized at some characteristic frequencies close, but not identical to the oscillator intrinsic frequency, and the optimized synchronization is attributed to the resonance between the periodic coupling and FLE. Our study sheds lights on the synchronization behavior of temporal networks, and provides insights on the interplay between the network dynamics and structure in evolutionary networks in general.

\section{Model and phenomenon}\label{sec1}

Our model of networked oscillators reads
\begin{equation}
\dot{\bm{x}}_i=\bm{F}_i(\bm{x}_i)+\varepsilon(t)\sum\limits^{N}_{j=1}a_{ij}[\bm{H}(\bm{x}_j)-\bm{H}(\bm{x}_i)],
\label{model}
\end{equation}
with
\begin{equation}
\varepsilon(t)=\varepsilon_0[1+\sin(\omega t)].
\label{coupling}
\end{equation}
Here, $i,j=1,2,\cdots,N$ are the oscillator (node) indices, $\bm{x}_i$ is the state vector of the $i$th oscillator, $\bm{F}_i$ describes the dynamics of the $i$th oscillator in the isolated form, and $\bm{H}$ represents the coupling function. 
%For simplicity, we set the oscillators to be of identical nodal dynamics. 
The coupling relationship of the oscillators is captured by the adjacency matrix $\bm{A}=\{a_{ij}\}$, with $a_{ij}=a_{ji}=1$ if there is a link between nodes $i$ and $j$ on the network, and $a_{ij}=0$ otherwise. The evolution of the network structure is reflected in the time-dependent coupling strength described by Eq. (\ref{coupling}), which is uniformly updated for all the network links. $\varepsilon_0$ and $\omega$ denote, respectively, the amplitude and frequency of the periodic coupling. This model, or its equivalent forms, describes the dynamics of a large variety of spatiotemporal systems, and has been employed as one of the standard models in exploring the synchronization behaviors of coupled oscillators~\cite{SYNREV:Boccaletti,SYNREV:Arenas}.

\begin{figure*}[tbp]
\begin{center}
\includegraphics[width=0.9\linewidth]{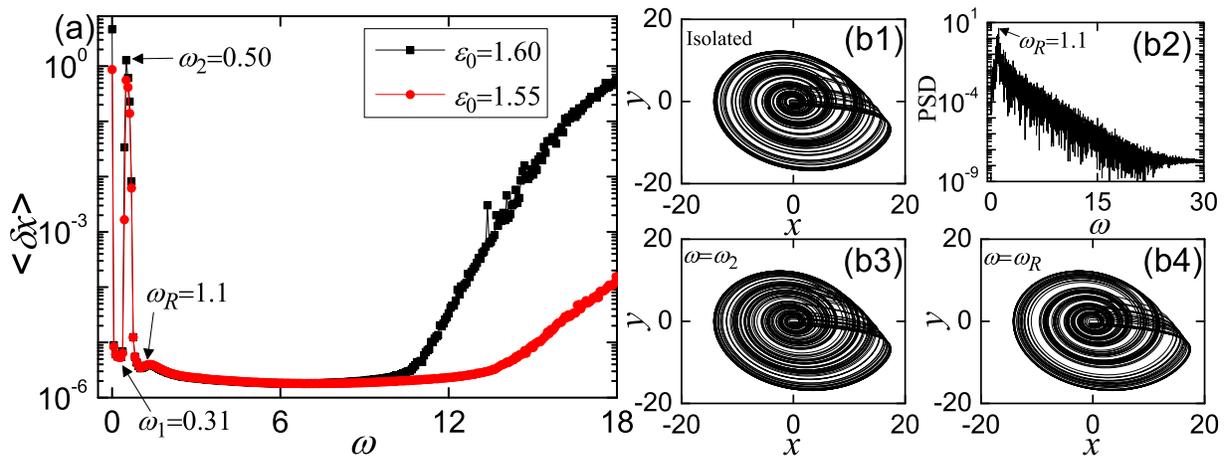}
\caption{(Color online) For three globally coupled chaotic R\"ossler oscillators, the influence of the coupling frequency, $\omega$, on network synchronization. (a) The variation of the averaged synchronization error, $\left<\delta x \right>$, with respect to $\omega$ for different coupling amplitudes, $\varepsilon_0=1.60$ and $1.55$. The results are averaged over a time period of $T=1 \times 10^3$ and $1000$ realizations. (b1) The trajectory of the $1$st oscillator in the isolated form in the $(x,y)$ plane. (b2) The PSD of the variable $x$ shown in (b1). The dominant component is locating at $\omega_R\approx 1.1$. (b3) By $\varepsilon_0=1.60$ and $\omega=\omega_2=0.5$, the trajectory of the $1$st oscillator. (b4) By $\varepsilon_0=1.60$ and $\omega=\omega_R$, the trajectory of the $1$st oscillator.}
\label{fig1}
\end{center}
\end{figure*}

To consolidate the model, we adopt the classical R\"ossler oscillator as the nodal dynamics and, for the sake of simplicity, employ the all-to-all coupling structure. The dynamics of isolated R\"ossler oscillator is described by equations $(dx/dt, dy/dt, dz/dt )^T = (-\Omega y-z, \Omega x+0.2y, 0.2+xz-9z)^T$~\cite{Rossler}. The characteristic phase frequency (or the intrinsic frequency) of the R\"ossler oscillator is governed by the parameter $\Omega$. In general, the larger is $\Omega$, the higher will be the intrinsic frequency, $\omega_R$. For $\Omega=1.0$, the oscillator presents the chaotic motion [see Fig. \ref{fig1}(b1)]. Numerically, the intrinsic frequency can be identified from the power spectral density (PSD) of the state variables. For $\Omega=1.0$, we have $\omega_R\approx 1.1$, as depicted in Fig. \ref{fig1}(b2). $\Omega$ characterizes the time scale of the nodal dynamics, around which the oscillating frequency of the coupling, $\omega$, will be tuned in our following study. Following Refs. \cite{CL:2009PRE,TIMME:PRL,AB:NODY}, we couple the oscillators through their $x$ variables, i.e., the coupling function is chosen as $\bm{H}([x,y,z]^T)=[x,0,0]^T$.

By the coupling amplitude $\varepsilon_0=1.60$, we plot in Fig. \ref{fig1}(a) the variation of the averaged synchronization error, $\left<\delta x\right>$, with respect to the coupling frequency, $\omega$, for $N=3$ globally coupled R\"{o}ssler oscillators. Here, the averaged synchronization error is defined as $\left<\delta x\right>\equiv\left<\sum_i |x_i-\bar{x}|/N\right>$, with $\bar{x}=\sum_i x_i/N$ the network-averaged state and $\left<\cdots\right>$ the time-average function. Apparently, the smaller is $\left<\delta x\right>$, the stronger is the network synchronized. In addition, to cope with the mismatch of parameters in realistic situations, we set the oscillators with different $\Omega$. Specifically, we set $\Omega=1.0$, $1.0+\delta\omega$ and $1.0-\delta\omega$ for oscillators $1$, $2$ and $3$, respectively, with $\delta\omega=1\times 10^{-3}$.  Figure \ref{fig1}(a) shows that, as $\omega$ increases from $0$ to $18$, the value of $\left<\delta x\right>$ is wildly changed. More specifically, when the coupling is static ($\omega=0$), we have $\left<\delta x\right>\approx 4.62$, indicating that the network is deeply desynchronized. However, as $\omega$ increases from $0$, the value of $\left<\delta x\right>$ is quickly decreased and, at about $\omega_1=0.31$, we have $\left<\delta x\right>\approx 0$, indicating that the network work is well synchronized at this point. Increasing $\omega$ further, it is shown that the value of $\left<\delta x\right>$ is gradually increased. After reaching its local maxima at about $\omega_2=0.50$, $\left<\delta x\right>$ begins to decrease again as $\omega$ increases. Then, in a wide range about $\omega\in(0.63,10.08)$, we have $\left<\delta x\right>\approx 0$. This forms the $2$nd window of synchronization in the parameter space of $\omega$. Finally, as $\omega$ exceeds $10.08$, the value of $\left<\delta x\right>$ is gradually increased. Clearly, the network synchronization is modified by tuning $\omega$. It is worth noting that in tuning $\omega$, the amplitude of the coupling strength, $\varepsilon_0$, is kept unchanged. As such, the tuning of $\omega$ provides actually a new approach for synchronization optimization.

To check whether the similar phenomenon is observable for other coupling amplitudes, we plot in Fig. \ref{fig1}(a) also the variation of $\left< \delta x \right>$ with respect to  $\omega$ for $\varepsilon_0=1.55$. It is seen that in the region of small $\omega$, the variation of $\left< \delta x \right>$ is very close to that of $\varepsilon_0=1.60$. However, in the region of large $\omega$, the value of $\left< \delta x \right>$ is clearly smaller to that of $\varepsilon_0=1.60$. In specific, for $\varepsilon_0=1.55$, the onset of network desynchronization occurs at $\omega\approx13.2$, while for $\varepsilon_0=1.60$ this occurs at $\omega\approx 10.08$. That is, the $2$nd synchronization window is enlarged by decreasing $\varepsilon_0$. Numerical results thus suggest that besides the coupling frequency, the synchronization performance is also influenced by the coupling amplitude.

In Ref.~\cite{AB:NODY}, it is reported that when $\omega\approx \omega_R$, the dynamics of the oscillators will be strongly affected by the periodic coupling, resulting in a desynchronization window in the parameter space of $\omega$. For our numerical results shown in Fig. \ref{fig1}(a), there does exist a desynchronization window, $\omega\in (0.48, 0.63)$, yet this window is away from $\omega_R$. On the contrary, as depicted in Fig. \ref{fig1}(a), the network is highly synchronized at $\omega_R$. The contradictory findings intrigue our interest of the impact of $\omega$ on the oscillator dynamics. We first check the dynamics of the oscillators for $\omega=\omega_2$, with which the network is mostly desynchronized. Figure \ref{fig1}(b3) shows the trajectory of the $1$st oscillator in the $(x,y)$ plane (the results for other oscillators are similar). It is seen that the trajectory is of no apparent difference to that of Fig. \ref{fig1}(b1). Meanwhile, the analysis of PSD shows that the dominant frequency is still locating at $\omega_{R}$. We next check the dynamics of the oscillators for $\omega=\omega_R$. The trajectory of the $1$st oscillator is plotted in Fig. \ref{fig1}(b4). Still, the dynamics of the oscillator is not apparently affected by the periodic coupling. Besides the special points $\omega_2$ and $\omega_R$, we have also checked the dynamics of the oscillators for other coupling frequencies (for $\varepsilon_0=1.55$ and $1.60$), and no apparent difference is found between the dynamics of coupled and isolated oscillators.

The verification of the oscillator dynamics is necessary and important. On the one hand, it excludes the possibility that the variation of network synchronization, as depicted in Fig. \ref{fig1}(a), is induced by the deformation of the oscillator trajectories. As such, the mechanism revealed in Ref.~\cite{AB:NODY} can not be used to explain the phenomenon observed here.  On the other hand, the observation that the trajectories of the oscillators are not  (or only slightly) affected by the periodic coupling makes it possible to conduct a theoretical analysis on the numerical results based on the MSF method. In the standard MSF method, the synchronous manifold is of the same dynamics to that of isolated oscillator. It is only under this condition that the curve of MSF is independent of the coupling strength and the network structure~\cite{MSF:Pecora,MSF:HG,MSF:Barahona}. As the calculation of the MSF relies on only the statistical properties of the synchronous manifold (i.e., the largest Lyapunov exponent), the unaffected trajectories under periodic coupling thus suggest that the phenomenon observed in simulations could be analyzed by a unique MSF, as will be detailed in the following section.

\section{Mechanism analysis}

\subsection{The MSF approach}

Although the MSF method is proposed for identical oscillators, it can be applied to oscillators of slight parameter mismatches as well~\cite{SJ:EPL,AS:EPL}. Treating the oscillators as identical, the MSF for periodically coupled oscillators can be obtained, as follows. Let $\bm{x}_s$ be the synchronous manifold of the oscillators and $\delta \bm{x}_i=\bm{x}_i-\bm{x}_s$ be an infinitesimal perturbation added to oscillator $i$, then whether the perturbed trajectories could be converged to the synchronous manifold is mainly determined by the set of variational equations
\begin{equation}
\delta \dot{\bm{x}}_i=\bm{DF}(\bm{x}_s)\delta \bm{x}_i+\varepsilon(t) \sum_{j=1}^N c_{ij}\bm{DH}(\bm{x}_s)\delta \bm{x}_j,
\label{vareq}
\end{equation}
with $i,j=1,\ldots,N$. Here, $\bm{C}$ is the coupling matrix, with $c_{ij}=a_{ij}$ and $c_{ii}=-k_i=-\sum_j a_{ij}$ for the non-diagonal and diagonal elements, respectively. $\bm{DF}(\bm{x}_s)$ and $\bm{DH}(\bm{x}_s)$ are the Jacobian matrices evaluated on $\bm{x}_s$. For the network to be synchronizable, the necessary condition is that $\delta \bm{x}_i$ decreases to $0$ with time for all the oscillators.

Transforming Eqs. (\ref{vareq}) into the mode space spanned by the eigenvectors of the coupling matrix, we have the set of decoupled variational equations
\begin{equation}
\delta \dot{\bm{y}}_i=\bm{DF}(\bm{x}_s)\delta \bm{y}_i+\varepsilon(t) \lambda_i \bm{DH}(\bm{x}_s)\delta \bm{y}_i,
\label{mode}
\end{equation}
with $\delta\bm{y}_i$ the $i$th mode, and $0=\lambda_1>\lambda_2\geq\lambda_3\geq \ldots\geq \lambda_N$ the eigenvalues of $\bm{C}$. The mode associated with $\lambda_1=0$ characterizes the motion in parallel to the synchronous manifold, whereas the other modes characterize the motion transverse to the synchronous manifold. In the mode space, the necessary condition for synchronization becomes that $\delta\bm{y}_i$ approaches 0 with time for all the transverse modes. Denote $\Lambda_i$ as the largest Lyapunov exponent of Eq. (\ref{mode}), this means that $\Lambda_i$ should be negative for $i=2,\ldots,N$. Introducing the generic coupling strength $\sigma\equiv -\varepsilon_0\lambda$, we have the generalized variational equation
\begin{equation}
\delta \dot{\bm{y}}=\bm{DF}(\bm{x}_s)\delta \bm{y}- \sigma[1+\sin(\omega t)] \bm{DH}(\bm{x}_s)\delta \bm{y},
\label{MSF}
\end{equation}
which stands as the MSF of our model. Solving Eq. (\ref{MSF}) numerically, we are able to obtain the variation of $\Lambda$ with respect to $\omega$ and $\sigma$, based on which the stable region of synchronization, i.e., the region with $\Lambda<0$, can be identified. Now, the condition for network synchronization becomes that $\Lambda_i<0$ for $i=2,\ldots,N$, i.e., all the transverse modes should be staying inside of the stable region in the two-dimensional parameter space $(\omega,\sigma)$.

\begin{figure*}[tbp]
\begin{center}
\includegraphics[width=0.95\linewidth]{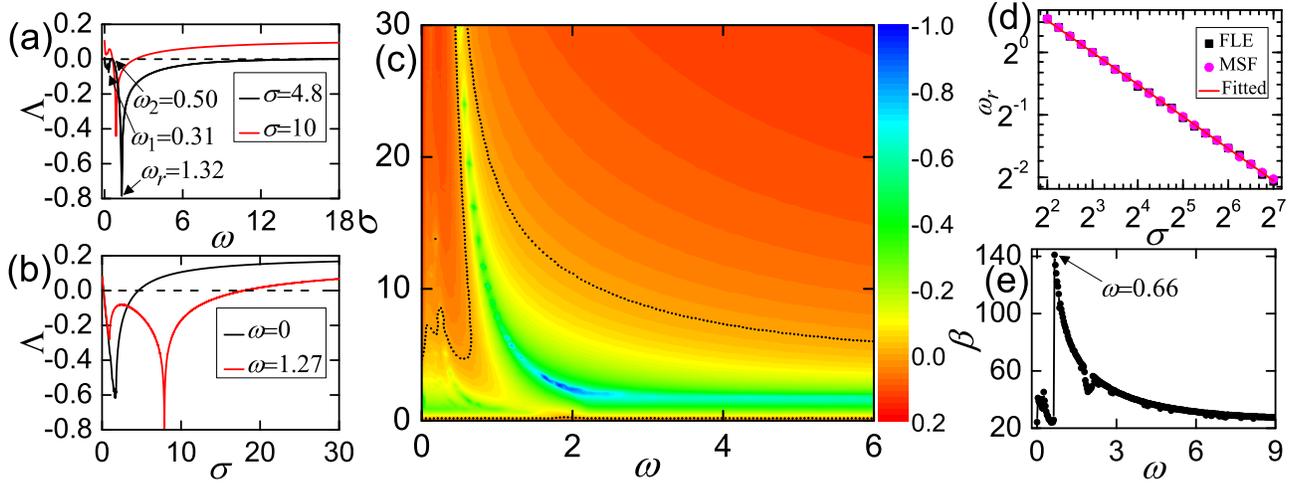}
\caption{(Color online) The results obtained by the MSF method. (a) The variation of $\Lambda$ with respect to $\omega$. For $\sigma=4.8$ ($\varepsilon_0=1.60$), $\Lambda$ is negative in the regions $\omega\in(0.06,0.44)$ and $\omega\in(0.63,10.08)$. Increasing $\sigma$ to $10$, the stable region is shrunk to $\omega\in(0.63,2.32)$. (b) The dependence of $\Lambda$ on $\sigma$. For static coupling ($\omega=0$), $\Lambda$ is negative in the region $\sigma\in(0.19,4.61)$. By periodic coupling of $\omega=1.27$, the stable region is enlarged to $\sigma\in(0.19, 17.97)$. (c) The contour plot of $\Lambda$ in the parameter space of $(\omega,\sigma)$. The boundaries of the stable region are marked by the dashed lines. (d) The variation of $\omega_r$, i.e., the characteristic frequency for optimized synchronization, with respect to $\sigma$. The fitted data gives $\omega_r\propto \sigma^{-\gamma}$, with $\gamma\approx 0.5$. Squares: the characteristic frequency, $\omega^f_r$, identified by the FLE method. (e) The variation of the network synchronizability, $\beta$, as a function of $\omega$. $\beta$ is diverged at $\omega\approx 0.66$.}
\label{fig2}
\end{center}
\end{figure*}

For the network studied in Fig. \ref{fig1}, we have $\lambda_2=\lambda_3=-3$, which, for $\varepsilon_0=1.60$, give $\sigma_2=\sigma_3=\varepsilon_0\lambda_{2,3}=4.8$. Fixing $\sigma=4.8$ in Eq. (\ref{MSF}), we plot in Fig. \ref{fig2}(a) the variation of $\Lambda$ as a function of $\omega$. It is seen that the variation of $\Lambda$ is in good agreement with the numerical results shown in Fig. \ref{fig1}(a). In particular, $\Lambda$ reaches its local minima around $\omega_1=0.31$, and reaches the local maxima around $\omega_2=0.5$ [where $\left<\delta x\right>$ reaches its local maxima in Fig. \ref{fig1}(a)]. Besides confirming the numerical observations, the MSF curve offers more information on the variation of network synchronization. For example, by a closer look at the behavior of $\Lambda$, it is found that the minimum $\Lambda$ is reached at $\omega_r\approx 1.32$, but not at $\omega_R=1.1$. We call $\omega_r$ the characteristic frequency for synchronization. To show the influence of $\sigma$ on MSF, we plot in Fig. \ref{fig2}(a) also the results for $\sigma=10$. Comparing to the case of $\sigma=4.8$, it is seen that the upper bound of the $2$nd synchronization window is significantly decreased (to about $\omega=1.3$). {\color{red}This result is in consistent with the phenomenon observed in Fig. \ref{fig1}(a), in which the $2$nd synchronization window is narrowed as $\varepsilon_0$ is increased from $1.55$ to $1.60$.} Moreover, Fig. \ref{fig2}(a) also shows that for $\sigma=10$, the location of the minimum $\Lambda$, i.e., the characteristic frequency, is shifted slightly to the left ($\omega_r\approx 1.3$).

Having justified the validity of the MSF method in quantifying the synchronizability of periodically coupled oscillators, we next employ the this method for a detailed analysis on the dependence of network synchronization on $\omega$ and $\sigma$. Figure \ref{fig2}(b) shows the variation of $\Lambda$ with respect to $\sigma$. For the case of static coupling ($\omega=0$), $\Lambda$ is negative in a bounded region $\sigma\in [0.19, 4.61]$. When periodic coupling with $\omega=1.27$ is adopted, the stable region is enlarged to $\sigma\in [0.19, 17.97]$. Denote $\sigma_l$ and $\sigma_u$ as the lower and upper bounds of the stable region, respectively, the network synchronizability then can be characterized by the ratio $\beta\equiv\sigma_u/\sigma_l$. For the given network structure, the larger is $\beta$, the wider will be the range for synchronization in the parameter space of $\omega$. We thus have $\beta=24.3$ for $\omega=0$ and $\beta=94.58$ for $\omega=1.27$, i.e., the network synchronizability is increased by about four times. Fig. \ref{fig2}(e) shows the detailed variation of $\beta$ with respect to $\omega$.

To have a global picture on the dependence of network synchronization on $\omega$ and $\sigma$, we plot in Fig. \ref{fig2}(c) the contour plot of $\Lambda$. Fig. \ref{fig2}(c) shows some interesting features overlooked in previous studies. Firstly, it is shown that by the periodic coupling, the upper bound of the stable region, $\sigma_u$, is significantly affected by varying $\omega$, but the lower bound, $\sigma_l$, is hardly changed. As such, by tuning $\omega$, it is the range of the stable region (or the network synchronizability), $[\sigma_l,\sigma_u]$ (or $\beta$), that is significantly enlarged, whereas the coupling cost (i.e., the smallest coupling amplitude for synchronization) is almost not affected. Secondly, the characteristic frequency where $\Lambda$ reaches its minima, $\omega_r$, in general is different from the oscillator intrinsic frequency, $\omega_R$, and, interestingly, is varying with $\sigma$. As depicted in Fig. \ref{fig2}(d), as $\sigma$ increases, $\omega_r$ is decreased by roughly a power-law scaling. This observation implies that the enhanced synchronization at $\omega_r$ is not induced by the phase-locking between the nodal dynamics and periodic coupling~\cite{AB:NODY}. Thirdly, the upper bound of the stable region is also varied violently at a small coupling frequency (around $\omega_1= 0.31$). This observation confirms again the irrelevance of phase-locking to synchronization enhancement, as this characteristic frequency is also decreased with $\sigma$. Finally, as $\sigma$ increases, the synchronization window is gradually narrowed. Meanwhile, with the increase of $\sigma$, the minimum value of $\Lambda$ is gradually increased. These observations indicate that the enhancement of synchronization by tuning $\omega$ is more prominent for small $\sigma$, which, in realistic situations, corresponds to complex networks of small eigenvalues and weak coupling strength.

\subsection{The FLE approach}

Whereas the influence of the coupling frequency on synchronization can be analyzed by the MSF method, the underlying mechanism is still not clear. In particular, it remains unknown why $\Lambda$ reaches its minima at $\omega_r$ and why $\omega_r$ is varying with $\sigma$. These questions call for a study on the temporal (local) stability of the MSF. Here we employ FLE for such a purpose~\cite{Pikovsky:Chaos1993,AP:PRE1999,FLE:XGW2006,KS:Chaos2010,AEB:2016}. The FLE is defined as
\begin{equation}
\Lambda^f(m)=\frac{1}{\Delta T}\ln|\bm{Q}_m(\Delta T)\cdot\bm{u}_0|,
\label{FLE}
\end{equation}
with $\Delta T$ the length of the time interval over which the FLE is averaged, $m$ the interval index, $\bm{Q}({\Delta T})$ the matrix solution of the equation $d\bm{Q}/dt=\bm{DF}(\bm{x}(\Delta T))\cdot\bm{Q}$, and $\bm{u}_0$ the random unit vector in the tangent space of the synchronous manifold. The trajectory is temporally (locally) stable if $\Lambda^f\leq 0$, otherwise it is temporally unstable.

For the typical chaotic motion, while the largest Lyapunov exponent is positive, the FLE could be negative for some time intervals. These intervals correspond to regions in the phase space in which infinitesimal vectors in fact contract in length ($\Lambda^f<0$). The asymptotic exponent, i.e., the largest Lyapunov exponent, is just the weighted sum of the temporally positive exponents when the trajectory visits the expanding regions ($\Lambda^f>0$) and the temporally negative exponents when the trajectory is in the contracting regions. For chaotic oscillator, the positive components weight over the negative ones, resulting in the positive asymptotic exponent~\cite{Pikovsky:Chaos1993,AP:PRE1999,FLE:XGW2006}.

\begin{figure*}[tbp]
\begin{center}
\includegraphics[width=0.8\linewidth]{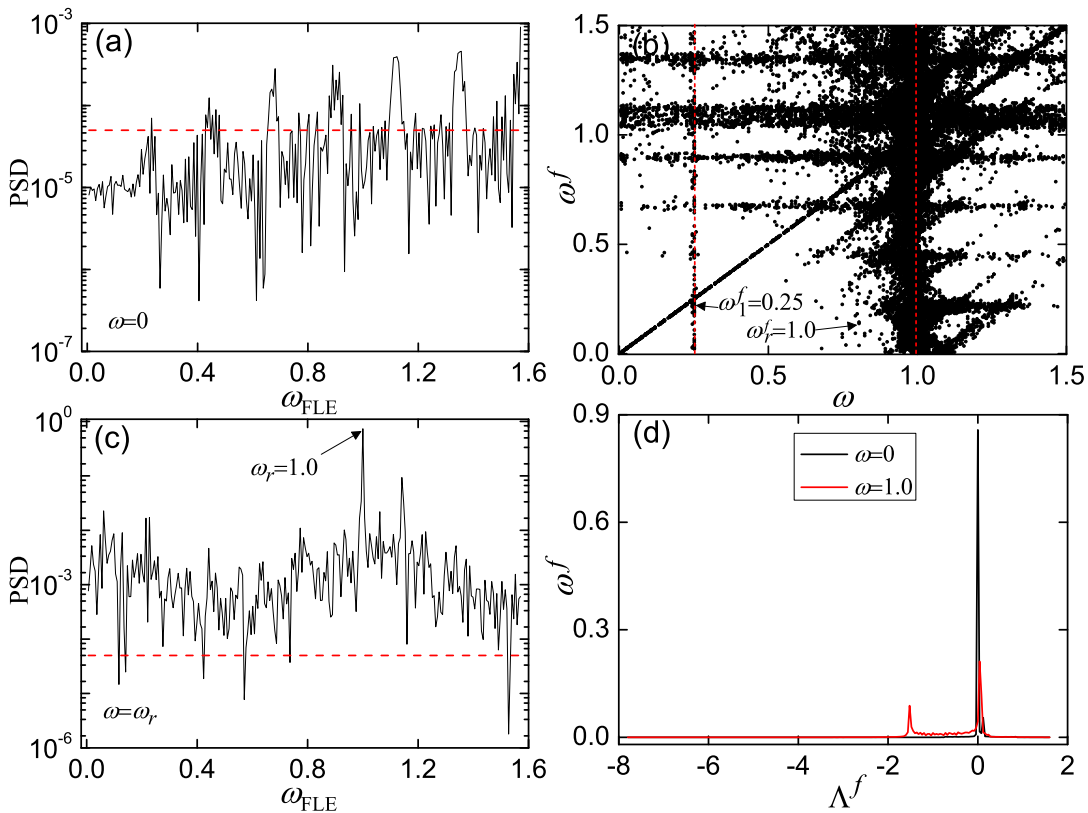}
\caption{(Color online) For $\sigma=8$, the results obtained by the method of FLE analysis. (a) By static coupling, the PSD of FLE. Dashed line denotes the threshold used to identify the dominant components. (b) The variation of the dominant PSD frequencies, $\omega^f$, with respect to the coupling frequency, $\omega$. The vertical dashed lines denote the characteristic frequencies, $\omega^f_1\approx 0.25$ and $\omega^f_r\approx 1$, where the dominant frequencies are broadly distributed. The diagonal line represents the component of the periodic coupling. (c) The PSD of FLE for $\omega=\omega^f_r$. (d) The probability distributions of FLE for $\omega=0$ and $\omega^f_r$.}
\label{fig3}
\end{center}
\end{figure*}

We proceed to study the response of FLE of the MSF to periodic coupling. In simulations, we fix $\sigma=8$ and set $\Delta T=0.2$, while noting that the findings to be reported in the following are independent of $\Delta T$ (given that $\Delta T$ is not too large). Fig. \ref{fig3}(a) shows the PSD of FLE for the case of static coupling, $\omega=0$. It is seen that the PSD is featured by the embedding of a few of dominant components over a broad spectral background. To focus on the dominant components, we truncate PSD by the threshold $5\times 10^{-5}$ and record only the dominant frequencies, $\omega^f$. We note that the dominant frequencies characterize the time scales of the local instability of the synchronous manifold, which is different from that of the nodal dynamics.

Figure \ref{fig3}(b) shows the variation of the truncated dominant frequencies with respect to $\omega$. Interestingly, it is found that the dominant frequencies are isolated from each other for most values of $\omega$, but are broadly distributed around two characteristic frequencies, $\omega^f_1\approx 0.25$ and $\omega^f_r=1.0$. Remarkably, these characteristic frequencies are very close to the characteristic frequencies, $\omega_{1,r}$, identified by the MSF method [see Fig. \ref{fig2}]. To have more details on the response of FLE at the characteristic frequencies, we plot in Fig. \ref{fig3}(c) the PSD of FLE for $\omega^f_r=1.0$. Comparing with the results of static coupling [Fig. \ref{fig3}(a)], it is seen that in Fig. \ref{fig3}(c) a large number of dominant frequencies are evoked, giving rise to the broad spectral distribution. As a consequence of this, the probability distribution of FLE is changed from the unimodal to bimodal distributions, with the new peak located around $\Lambda^f=-1.52$ [Fig. \ref{fig3}(d)]. It is just the appearance of this new peak that leads to the sharp decrease of $\Lambda$ in the MSF curve. The similar phenomenon is also observed around $\omega^f_1$, where many components are evoked in the PSD of FLE [see Fig. \ref{fig3}(b)] and $\Lambda$ reaches its local mimima in the MSF curve [Fig. \ref{fig2}(a)].

As for the results of MSF analysis, the characteristic frequencies, $\omega^f_{1,r}$, identified by the FLE method are also dependent of the generic coupling strength, $\sigma$. By the same procedure described above, we calculate $\omega^f_r$ for different $\sigma$. The results are also presented in Fig. \ref{fig2}(d). It is seen that $\omega^f_r$ falls exactly on the line fitted by $\omega_r$. The excellent agreement between $\omega^f_r$ and $\omega_r$ manifests the appropriateness of the FLE approach in exploring the synchronization phenomenon of periodically coupled oscillators, and, more importantly, points out the fact that the enhanced synchronization at the characteristic frequencies is due to the resonance between the periodic coupling and the temporal instability of the synchronous manifold (instead of the nodal dynamics).

\subsection{Other oscillators}

Comparing the FLE spectra of $\omega^f_1$ and $\omega^f_r$ in Fig. \ref{fig3}(b), it is seen that more components are evoked by $\omega^f_r$ than $\omega^f_1$. Meanwhile, in Fig. \ref{fig2}(a) it is shown that comparing to $\omega_1^f$, the value of $\Lambda$ is smaller at $\omega^f_r$. This observation leads to our following hypothesis: the stronger is the resonance between FLE and the periodic coupling, the more significant will be the network synchronization enhanced. We next employ the model of coupled Lorenz oscillators to verify this hypothesis. The Lorenz oscillator in the isolated form is described by equations $(dx/dt, dy/dt, dz/dt )^T = (10y-10x, 28x-xz-y, xy-8z/3)^T$, which presents the chaotic motion in the phase space~\cite{Lorenz}. The oscillators are coupled by the function $\bm{H}([x,y,z]^T)=[0,x,0]^T$, i.e., the $y$ component is coupled to the $x$ component. Still, the coupling strength is varying with time periodically, as described by Eq. (\ref{coupling}).

\begin{figure}[tbp]
\begin{center}
\includegraphics[width=0.85\linewidth]{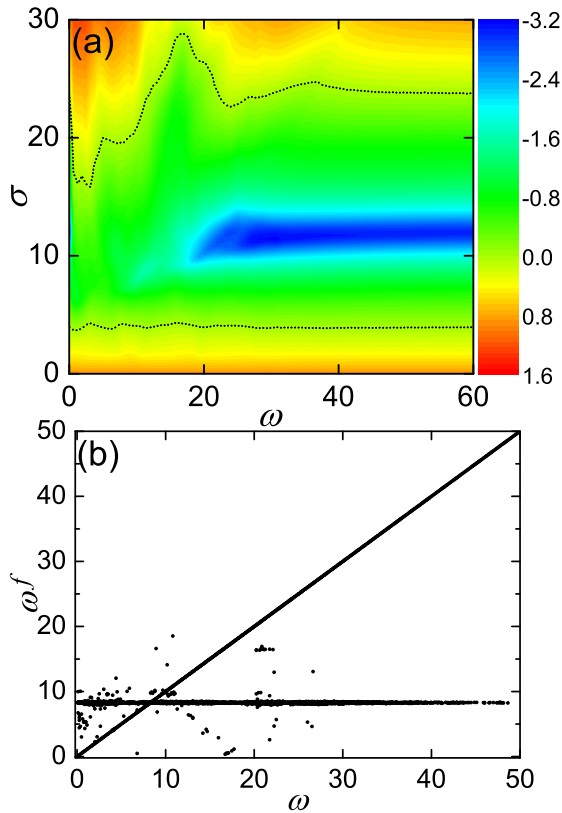}
\caption{(Color online) The results for chaotic Lorenz oscillators. (a) The variation of $\Lambda$ with respect to $\omega$ and $\sigma$. (b) Fixing $\sigma=11$, the variation of the dominant frequencies, as identified from the PSD of FLE, with respect to $\omega$. The diagonal line corresponds to the frequency of periodic coupling.}
\label{fig4}
\end{center}
\end{figure}

By solving Eq. (\ref{MSF}) numerically, we plot in Fig. \ref{fig4}(a) the contour plot of $\Lambda$ in the two-dimensional parameter space $(\omega,\sigma)$. It is seen that as $\omega$ varies, the range of the stable region, $(\sigma_l,\sigma_u)$, is also modulated. However, comparing to the results of R\"ossler oscillators (Fig. \ref{fig2}), the influence of $\omega$ on synchronization is less prominent. Fixing $\sigma=11$ (where $\sigma_u$ is maximized), we plot in Fig. \ref{fig4}(b) the variation of the dominant frequencies of FLE with respect to $\omega$. Here, $\Delta T=6\times 10^{-2}$ and the truncation threshold is chosen as $2\times 10^{-2}$ (the results for other parameters are qualitatively the same). It is seen that, unlike the case of R\"ossler oscillators [Fig. \ref{fig3}(b)], there is no characteristic frequency by which the distribution of the dominant frequencies is violently changed.

\section{Discussions and conclusion}

The finding that network synchronization is enhanced by periodic coupling provides an alternative approach for synchronization optimization. According to the MSF paradigm, the synchronizability of networked oscillators is jointly determined by two factors: the shape of the MSF curve and the eigenvalues of the network coupling matrix. In conventional studies where constant coupling is adopted, the stable region of the MSF is determined solely by the nodal dynamics and coupling function. As such, to improve the network synchronizability, the only available approach is to modify the eigenvalues of the network coupling matrix. This has led to the extensive studies on the impact of network structure on synchronization over the past two decades~\cite{SYNREV:Boccaletti,SYNREV:Arenas}. The present work, however, adopts a different approach for optimizing synchronization: modifying the MSF curve. By the periodic coupling, the MSF contains two independent parameters, $\sigma$ and $\omega$. As demonstrated in Fig. \ref{fig2}, the stable region of the MSF can be effectively modified by changing $\omega$. In particular, at some characteristic frequencies (e.g., $\omega_r$), the upper bound of the stable region can be significantly increased, making the stable region of MSF significantly enlarged. For network of fixed coupling matrix (i.e., the eigenvalues are fixed), the larger is the stable region, the higher will be the propensity for network synchronization. As a result of the enlarged stable region, the network synchronization is optimized at the characteristic frequencies.

A few remarks on the impact of periodic coupling on network synchronization are in order. First, the enlarged stable region of MSF is mainly attributed to the increase of the upper bound, $\sigma_u$. By varying the coupling frequency, the lower bound, $\sigma_l$, is only slightly changed [see Fig. \ref{fig2}]. This restricts the phenomenon of periodic-coupling-enhanced synchronization to be only observable for MSF of bounded stable region. For MSF of semi-open stable region, i.e., only $\sigma_l$ exists, the impact of periodic coupling on synchronization will be less prominent. Second, the enhanced synchronization at the characteristic frequencies, $\omega_{1,r}$, is rooted in the nonlinear response of the temporal instability of the synchronous manifold, i.e., the FLE, to the periodic driving. To realize the nonlinear response, a necessary condition is that the coupling matrix must be non-diagonal, i.e., $\bm{H}([x,y,z]^T)\neq [x,y,z]^T$. If the coupling matrix is diagonal, the stable region will be not affected by the periodic coupling, even when the time scales of the periodic coupling and the nodal dynamics are comparable. Third, whereas our study points out the connection between the periodic coupling and the temporal stability of the synchronous manifold, further studies are still needed to understand the mechanism of enhanced synchronization by periodic coupling. As a simple measure of the temporal stability, we employ FLE in the present work to investigate the response of MSF to periodic coupling. Although violent changes are observed in FLE at the characteristic frequencies, it remains unknown to us how this happens. It is our hope that this question could be addressed in the near future by examining more examples and employing new mathematical techniques.

To summarize, we have studied the synchronization behavior of networked chaotic oscillators with periodic coupling, and found that the network synchronization can be effectively enhanced by tuning the coupling frequency. Based on the method of MSF, we have conducted a detailed analysis on the impact of the coupling frequency on synchronization, and found that network synchronization is optimized at some characteristic frequencies comparable to that of the nodal dynamics. The mechanism of synchronization optimization is investigated by the technique of FLE, and it is revealed that the optimized synchronization is due to a violent change of the PSD of FLE at the characteristic frequencies. Our study sheds new lights on the synchronization behavior of temporally coupled oscillators, and the findings might have applications to the optimization of synchronization in realistic networks.

This work was supported by the National Natural Science Foundation of China under the Grant Nos.~11375109 and 61703257, and by the Fundamental Research Funds for the Central Universities under the Grant No.~GK201601001.


\begin{thebibliography}{99}
%Review
\bibitem{SYNBOOK:Kuramoto} Y. Kuramoto, \emph{Chemical Oscillations, Waves, and Turbulence} (Springer, Berlin, 1984).

\bibitem{SYNBOOK:Pikovsky} A. Pikovsky, M. Rosenblum, and J. Kurths, \emph{Synchronization: A Universal Concept in Nonlinear Science} (Cambridge University Press, Cambridge, 2003).

\bibitem{SYNBOOK:Strogatz} S. H. Strogatz, \emph{Sync: The Emerging Science of Spontaneous Order} (Hyperion, New York, 2003).

\bibitem{SYNREV:Boccaletti} S. Boccaletti, V. Latora, Y. Moreno, M. Chavez, and D.-U. Hwang, Complex networks: Structure and dynamics, Phys. Rep. {\bf 424}, 175 (2006).

\bibitem{SYNREV:Arenas} A. Arenas, A. D\'iaz-Guilera, J. Kurths, Y. Moreno, and C. S. Zhou, Synchronization in complex networks, Phys. Rep. {\bf 469}, 93 (2008).

%MSF
\bibitem{MSF:Pecora} L. M. Pecora and T. L. Carroll, Master stability functions for synchronized coupled systems, Phys. Rev. Lett. {\bf 80}, 2109 (1998).

\bibitem{MSF:HG} G. Hu, J. Z. Yang, and W. J. Liu, Instability and controllability of linearly coupled oscillators: Eigenvalue analysis, Phys. Rev. E {\bf 58}, 4440 (1998).

\bibitem{MSF:Barahona} M. Barahona and L. M. Pecora, Synchronization in small-world systems, Phys. Rev. Lett. {\bf 89}, 054101 (2002).

% Net Syn
\bibitem{NETSYN:CKHU} P. M. Gade and C.-K. Hu, Synchronous chaos in coupled map lattices with small-world interactions, Phys. Rev. E {\bf 62}, 6409 (2000).

\bibitem{NETSYN:Nishikawa} T. Nishikawa, A. E. Motter, Y.-C. Lai, and F. C. Hoppensteadt, Heterogeneity in oscillator networks: Are smaller worlds easier to synchronize? Phys. Rev. Lett. {\bf 91}, 014101 (2003).

\bibitem{NETSYN:Motter2005} A. E. Motter, C. Zhou, and J. Kurths, Weighted networks are more synchronizable: How and why, AIP Conf. Proc. {\bf 776}, 201 (2005).

\bibitem{NETSYN:WXG2007} X. G. Wang,Y.-C.Lai, and C.-H. Lai, Enhancing synchronization based on complex gradient networks, Phys. Rev. E {\bf 75}, 056205 (2007).

%temporal network
\bibitem{TemNet:Holme} P. Holme and J. Saram\"aki, Temporal networks, Phys. Rep. {\bf 519}, 97 (2012).

\bibitem{TemNet:LiAming} A. Li, S. P. Cornelius, Y.-Y. Liu, L. Wang, and A.-L. Barab\'asi, The fundamental advantages of temporal networks, Science \textbf{358}, 1042 (2017).

%evolutionary network
\bibitem{SND:2002} S. N. Dorogovtsev and J. F. F. Mendes, Evolution of networks, Adv. Phys. \textbf{51}, 1079 (2002).

\bibitem{SG:2007}G. Szab\'o and G. F\'ath, Evolutionary games on graphs, Phys. Rep. {\bf 446}, 97 (2007).

\bibitem{ZCS:2006} C. Zhou, J. Kurths, Dynamical weights and enhanced synchronization in adaptive complex networks, Phys. Rev. Lett. \textbf{96}, 164102 (2006).

\bibitem{FW:2011} C. Fu and X. G. Wang, Network growth under the constraint of synchronization stability, Phys. Rev. E \textbf{83}, 066101 (2011).

\bibitem{MWFDL:2011} M. Li, X. G. Wang, Y. Fan, Z. Di, and C.-H. Lai, Onset of synchronization in weighted complex networks: The effect of weight-degree correlation, Chaos \textbf{21}, 025108 (2011).

\bibitem{MOTTER:2005} A. E. Motter and Y.-C. Lai, Cascade-based attacks on complex networks, Phys. Rev. E {\bf 66}, 065102 (2002).

\bibitem{Holme:2006} P. Holme and M. E. J. Newman, Nonequilibrium phase transition in the coevolution of networks and opinions, Phys. Rev. E \textbf{74}, 056108 (2006).

\bibitem{IBS:2010} I. B. Schwartz and L. B. Shaw, Rewiring for adaptation, Physics {\bf 3}, 17 (2010).

\bibitem{WYF:2016} Y. F. Wang, H. W. Fan, W. J. Lin, Y.-C. Lai, and X. G. Wang, Growth, collapse, and self-organized criticality in complex networks, Sci. Rep. {\bf 6}, 24445 (2016).

%time-dependent coupling
\bibitem{Belykh:2004} I. V. Belykh, V. N. Belykh, and M. Hasler, Blinking model and synchronization in small-world networks with a time-varying coupling, Physica D {\bf 195}, 188 (2004).

\bibitem{Boccaletti:2006} S. Boccaletti, D.-U. Hwang, M. Chavez, A. Amann, J. Kurths, and L. M. Pecora, Synchronization in dynamical networks: Evolution along commutative graphs, Phys. Rev. E {\bf 74}, 016102 (2006).

\bibitem{CM:2007} M. Chen, Synchronization in time-varying networks: A matrix measure approach, Phys. Rev. E {\bf 76}, 016104 (2007).

\bibitem{SO:2008}F. Sorrentino and E. Ott, Adaptive synchronization of dynamics on evolving complex networks, Phys. Rev. Lett. \textbf{100}, 114101 (2008).

\bibitem{FM:2008PRL} M. Frasca, A. Buscarino, A. Rizzo, L. Fortuna, and S. Boccaletti, Synchronization of moving chaotic agents, Phys. Rev. Lett. {\bf 100}, 044102 (2008).

\bibitem{CL:2009PRE} L. Chen, C. Qiu, and H. B. Huang, Synchronization with on-off coupling: Role of time scales in network dynamics, Phys. Rev. E {\bf 79}, 045101 (2009).

\bibitem{CL:2010EPJB} L. Chen, C. Qiu, H. B. Huang, G. X. Qi, and H. J. Wang, Facilitated synchronization of complex networks through a discontinuous coupling strategy, Eur. Phys. J. B {\bf 76}, 625 (2010).

\bibitem{Porfiri:2012} M. Porfiri, Stochastic synchronization in blinking networks of chaotic maps, Phys. Rev. E {\bf 85}, 056114 (2012).

\bibitem{Hasler:SIADS2} M. Hasler, V. N. Belykh, and I. V. Belykh, Dynamics of stochastically blinking systems, Part I: Finite time properties, SIAM J. Appl. Dyn. Syst. {\bf 12}, 1007 (2013).

\bibitem{VK:2014} V. Kohar, P. Ji, A. Choudhary, S. Sinha and J. Kurths, Synchronization in time-varying networks, Phys. Rev. E {\bf 90}, 022812 (2014).

\bibitem{TIMME:PRL} M. Schr\"oder, M. Mannattil, D. Dutta, S. Chakraborty and M. Timme, Transient uncoupling induces synchronization, Phys. Rev. Lett. {\bf 115}, 054101 (2015).

\bibitem{ZhouJ:2016} J. Zhou, Y. Zou, S. G. Guan, Z. H. Liu, and S. Boccaletti, Synchronization in slowly switching networks of coupled oscillators, Sci. Rep. {\bf 6}, 35979 (2016).

\bibitem{FN:2016Chaos} N. Fujiwara, J. Kurths, and A. Di\'az-Guilera, Synchronization of mobile chaotic oscillator networks, Chaos {\bf 26}, 094824 (2016).

\bibitem{AB:NODY} A. Buscarino, M. Frasca, M. Branciforte, L. Fortuna, J. C. Sprott, Synchronization of two R\"ossler systems with switching coupling, Nonlinear Dyn. {\bf 88}, 673 (2017).

\bibitem{Golovneva:2017} O. Golovneva, R. Jeter, I. Belykh, M. Porfiri, Windows of opportunity for synchronization in stochastically coupled maps, Physica D {\bf 340}, 1 (2017).

%others

\bibitem{Rossler} O. E. R\"ossler, An equation for continuous chaos, Phys. Lett. A {\bf 57}, 397 (1976).

\bibitem{SJ:EPL} J. Sun, E. M. Bollt, and T. Nishikawa, Master stability functions for coupled nearly identical dynamical systems, EPL {\bf 85}, 60011 (2009).

\bibitem{AS:EPL} S. Acharyya and R. E. Amritkar, Synchronization of coupled nonidentical dynamical systems, EPL {\bf 99}, 40005 (2012).

%FLE
\bibitem{Pikovsky:Chaos1993} A. Pikovsky, Local Lyapunov exponents for spatiotemporal chaos, Chaos {\bf 3}, 225 (1993).

\bibitem{AP:PRE1999} A. Prasad and R. Ramaswamy, Characteristic distributions of finite-time Lyapunov exponents, Phys. Rev. E {\bf 60}, 2761 (1999).

\bibitem{FLE:XGW2006} X. G. Wang, Y.-C. Lai, and C.-H. Lai, Characterization of noise-induced strange nonchaotic attractors, Phys. Rev. E {\bf 74}, 016203 (2006).

\bibitem{KS:Chaos2010} K. Stefa\'nski, K. Buszko, and K. Piecyk, Transient chaos measurements using finite-time Lyapunov exponents, Chaos {\bf 20}, 033117 (2010).

\bibitem{AEB:2016} A. E. Botha, Characteristic distribution of finite-time Lyapunov exponents for chimera states, Sci. Rep. {\bf 6}, 29213 (2016).

\bibitem{Lorenz} E. N. Lorenz, Deterministic nonperiodic flow, J. Atmos. Sci. {\bf 20}, 130 (1963).

\end{thebibliography}
\end{document}